\documentclass[final]{aipproc}
\layoutstyle{6x9}
\usepackage{epsfig}
\usepackage{rotate}

\newcommand{\insertfig}[2]{\mbox{\epsfxsize=#1cm \epsfbox{#2.eps}}}
\newcommand{\ft}[2]{{\textstyle\frac{#1}{#2}}}
\newcommand{\Bx}{x_{\rm B}}

\begin{document}

\begin{titlepage}

\begin{flushright}
DOE/ER/40762-289 \\ [-1mm] UMD-PP\#03-061
\end{flushright}

\centerline{\bf Renormalons in exclusive meson electroproduction}

\vspace{15mm}

\centerline{\bf A.V. Belitsky}

\vspace{15mm}

\centerline{\it Department of Physics}
\centerline{\it University of Maryland at College Park}
\centerline{\it College Park, MD 20742-4111, USA}

\vspace{15mm}

\centerline{\bf Abstract}

\vspace{0.5cm}

\noindent
We discuss the possibility of measuring generalized parton distributions in
exclusive electroproduction of mesons off the nucleon and estimate the
uncertainty from perturbatively induced higher-twist corrections. We find
that, while the magnitude of the cross section changes significantly taking
into account twist-four contributions modeled via renormalons, the transverse
spin asymmetry is weakly sensitive to them and displays the precocious
scaling.

\vspace{80mm}

\centerline{\it An extraction from talks given at}
\centerline{\it CIPANP 2003 (New York City, May 2003)}
\centerline{\it Symposium ``A celebration of JLab physics" (Jefferson Lab, June 2003)}
\centerline{\it Annual meeting of American Physical Society (Philadelphia, April 2003)}

\setcounter{page}{0}

\end{titlepage}

\title{Renormalons in exclusive meson electroproduction}

\author{A.V. Belitsky}{
address={Department of Physics, University of Maryland \\
         College Park, MD 20742-4111, USA}
}

\begin{abstract}

We discuss the possibility of measuring generalized parton distributions in
exclusive electroproduction of mesons off the nucleon and estimate the
uncertainty from perturbatively induced higher-twist corrections. We find
that, while the magnitude of the cross section changes significantly taking
into account twist-four contributions modeled via renormalons, the transverse
spin asymmetry is weakly sensitive to them and displays the precocious
scaling.

\end{abstract}

\maketitle


Generalized parton distributions (GPDs) $F (x, \xi, \Delta^2)$ encode exhaustive
information on one-particle correlations in the nucleon and thus carry the lore on
its wave function and the phase structure of the latter. The quantum-mechanical wave
function ${\mit\Psi}$ allows to predict expectation values of all observable for a
given system. An identical description is achieved by means of the density matrix
$\rho (x_1, x_2) = {\mit\Psi}^\ast (x_1) {\mit\Psi} (x_2)$. The latter can be used
in turn to construct the quantum equivalent of the classical phase-space distribution,
the primary example being known as the Wigner quasi-probability function $W (k, r) =
\int \frac{d x}{2 \pi \hbar} {\rm e}^{- i k x/\hbar} \rho (r - \ft12 x, r + \ft12 x)$.
Contrary to its classical counterpart it is not positive definite, --- a hallmark
of the interference. The marginals of $W (k, r)$ acquire however the probability
interpretation as coordinate density $\rho (r) = \int d k W (k, r) = |{\mit\Psi}
(r)|^2$, or equivalently the Fourier transform of the atomic form factor, and
momentum-space distribution $n (k) = \int \frac{d r}{2 \pi \hbar} W(k, r) =
|\widetilde {\mit\Psi} (k)|^2$ with $\widetilde {\mit\Psi} (k) = \int \frac{d x}{2
\pi \hbar} {\rm e}^{- i k x/\hbar} {\mit\Psi} (x)$. The $W (k, r)$ is an analogue
of a Fourier transformed one-dimensional GPD. The impact parameter-dependent parton
distributions \cite{Sop77}, related to GPDs again by a Fourier transform with respect
to the momentum transfer $\Delta_\perp$, are transparently identified as relativistic
nucleon Wigner distributions \cite{Bel03}. Thus, the studies of GPDs will shed the
light on the phase-space distribution of quarks in the proton.
\begin{figure}[t]
\mbox{
\begin{picture}(0,180)(208,0)
\put(0,0){\insertfig{8}{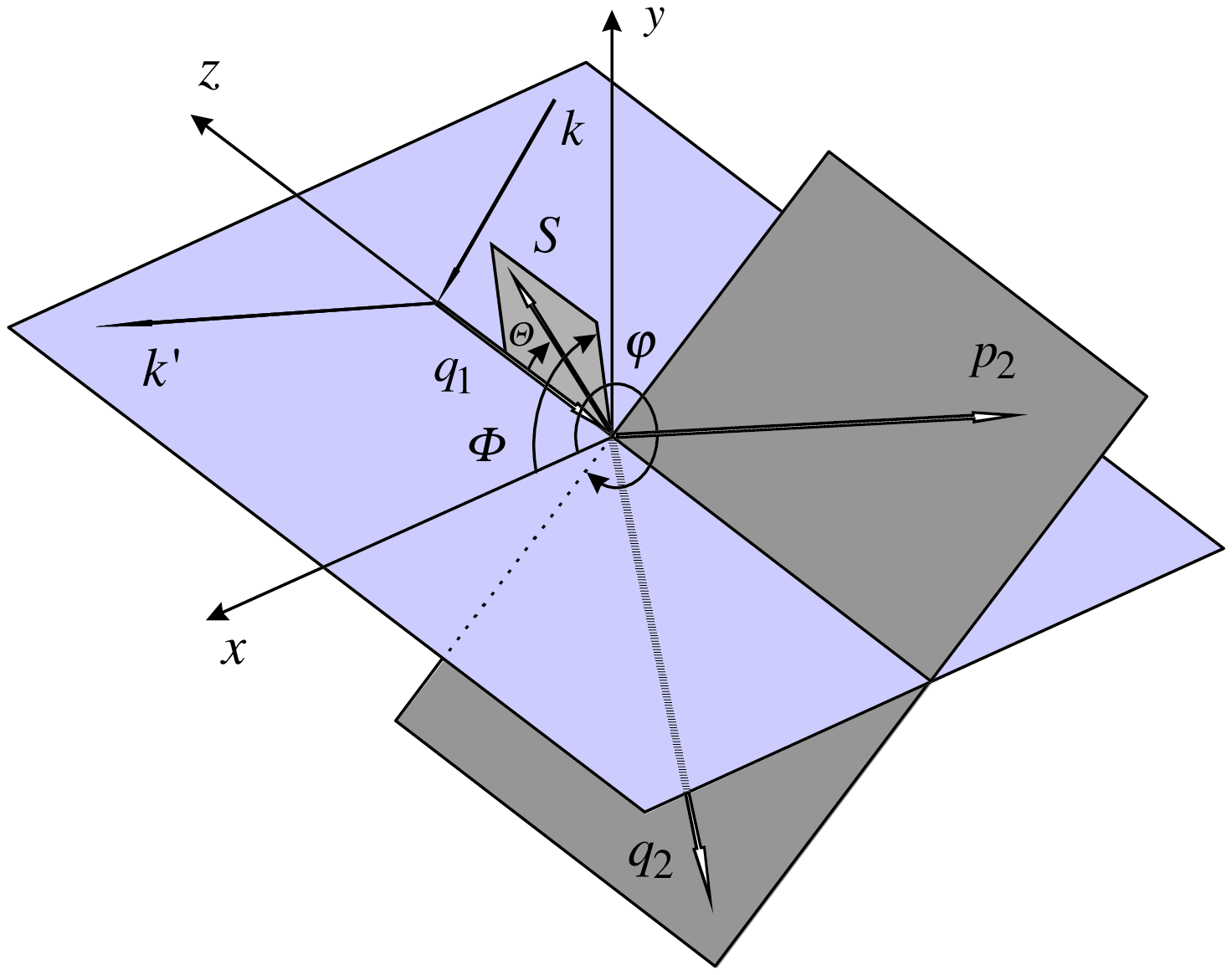}}
\put(260,5){\insertfig{5.4}{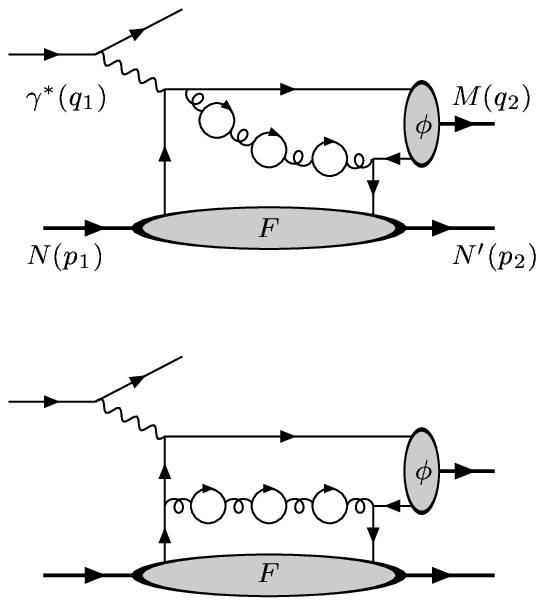}}
\end{picture}
}
\caption{\label{KinamticsAndDiagrams} Kinematics of the exclusive meson
electroproduction off the proton in its rest frame (left) and leading order
perturbative diagrams in hard scattering approach (right) dressed by fermion
bubble insertions which generate power corrections in the amplitude.}
\end{figure}
GPDs are cleanly probed in deeply virtual Compton scattering involving only one
hadron, the nucleon, whose structure is unraveled through electron scattering
\cite{BelMulKir01}. The same GPDs enter the amplitude of exclusive electroproduction
of mesons in the asymptotic regime of large momentum transfer \cite{ColFraStr97}.
However due to the presence of an extra hadron in the final state and the specifics
of perturbative QCD approach to such processes, one has to pose the question of
applicability of hard gluon exchange mechanism at moderate photon virtualities.
The cross section of meson photoproduction with longitudinally polarized $\gamma^\ast$
is, see Fig.\ \ref{KinamticsAndDiagrams} (left),
\begin{eqnarray}
\label{XsectionFinal}
\frac{d \sigma_{\scriptscriptstyle \! L}^{M}}{d |\Delta^2| d \varphi}
=
\frac{\alpha_{\rm em} \pi}{{\cal Q}^6}
\frac{f_M^2}{N_c^2} \frac{\Bx^2}{(2 - \Bx)^2}
\left\{
\sigma_M + \sigma^\perp_M \sin{\mit\Theta} \sin ({\mit\Phi} - \varphi)
\right\}
\, ,
\end{eqnarray}
where $q_1^2 = - {\cal Q}^2$, $x_{\rm B} = {\cal Q}^2/(2 p_1 \cdot q_1)$, and
$M$ runs over mesons of different flavors. For the charged psedoscalar meson
$P = \pi^+$ the decay constant is $f_\pi = 132 \, {\rm MeV}$, while for the
neutral vector mesons $V^0 = \rho^0, \omega$, they are $f_\rho = 153 \, {\rm MeV}$
and $f_\omega = 138 \, {\rm MeV}$. The parts of the cross sections read for
target polarization-independent \cite{FraPobPolStr98,BelMul01,GoePolVan01}
\begin{eqnarray}
\sigma_P
\!\!\!&=&\!\!\!
8 (1 - \Bx) | \widetilde {\cal H}_P |^2
-
\Bx^2 \frac{\Delta^2}{2 M_N^2}
| \widetilde {\cal E}_P |^2
-
4 \Bx^2 \, \Re{\rm e}
\left(
\widetilde {\cal H}^\ast_P \widetilde {\cal E}_P
\right)
\, , \\
\sigma_V
\!\!\!&=&\!\!\!
8 (1 - \Bx) | {\cal H}_V |^2
-
\Bx^2
\left(
2
+
(2 - \Bx)^2 \frac{\Delta^2}{2 M_N^2}
\right)
| {\cal E}_V |^2
-
4 \Bx^2 \, \Re{\rm e}
\left(
{\cal H}^\ast_V {\cal E}_V
\right)
\, ,
\end{eqnarray}
and target (transverse) polarization-dependent components
\begin{eqnarray}
\sigma^\perp_P
\!\!\!&=&\!\!\!
-
4 \Bx \sqrt{1 - \Bx} \sqrt{- \frac{\Delta^2}{M_N^2}}
\sqrt{1 - \frac{\Delta^2_{\rm min}}{\Delta^2}} \,
\Im{\rm m}
\left(
\widetilde {\cal H}^\ast_P \widetilde {\cal E}_P
\right)
\, , \\
\sigma^\perp_V
\!\!\!&=&\!\!\!
4 (2 - \Bx) \sqrt{1 - \Bx} \sqrt{- \frac{\Delta^2}{M_N^2}}
\sqrt{1 - \frac{\Delta^2_{\rm min}}{\Delta^2}} \,
\Im{\rm m}
\left(
{\cal H}^{\ast}_V {\cal E}_V
\right)
\, ,
\end{eqnarray}
respectively. The generalized structure function ${\cal F} = \{ {\cal H},
{\cal E}, \widetilde {\cal H}, \widetilde {\cal E}\}$ depends on the
skewness $\xi = x_{\rm B}/(2 - x_{\rm B})$, the $t$-channel momentum transfer
$\Delta^2$ and resolution scale ${\cal Q}^2$. In leading-twist approximation,
it is expressed as a convolution of the meson distribution amplitude $\phi (u)$,
normalized to $\int_0^1 d u \phi (u) = 1$, the quark or gluon GPD $F = \{ H, E,
\widetilde H, \widetilde E \}$ and, correspondingly, the quark or gluon
coefficient function $T$ via \cite{ColFraStr97}
\begin{eqnarray}
&&{\cal F}_M (\xi, \Delta^2; {\cal Q}^2)
\\
&&\qquad\equiv
\int_0^1 d u \int_{- 1}^1 d x \, \phi_M (u)
\left\{
T_M (u, x, \xi; {\cal Q}^2) F_M (x, \xi, \Delta^2)
+
T_g (u, x, \xi) F_g (x, \xi, \Delta^2)
\right\}
\, . \nonumber
\end{eqnarray}
The function $T_M$ to lowest order approximation is given by the one-gluon
exchange mechanism diplayed in Fig.\ \ref{KinamticsAndDiagrams} (right). The
studies of higher-order perturbative corrections to the hard coefficient function
in many physical observables have demonstrated that ambiguities generated by the
perturbative resummation of fermion vacuum polarization insertions were of the
same order of magnitude as available non-perturbative estimates of matrix elements
of higher-twist operators. The development and sophistication of these ideas has
led to some evidence that infrared renormalons might reflect the magnitude of
higher-twist contributions and even their functional dependence on scaling
variables and can thus be used as a rough estimate of power-suppressed effects.
On the practical side to compute them in the present circumstances, one replaces
the tree gluon propagator, in the single bubble-chain approximation, see Fig.\
\ref{KinamticsAndDiagrams} (right), by [in the Landau gauge]
\begin{eqnarray*}
{\cal D}_{\mu\nu} (k) = \frac{4 \pi}{\alpha_s b} \int_0^\infty d \tau \,
{\rm e}^{- 4 \pi/(\alpha_s b) \tau}
\left( \frac{\mu^2 {\rm e}^C}{- k^2} \right)^\tau
\frac{1}{k^2} \left( g_{\mu\nu} - \frac{k_\mu k_\nu}{k^2}\right)
\, ,
\end{eqnarray*}
where $C_{\overline{\rm\scriptscriptstyle MS}} = \ft53$ in the $\overline{\rm MS}$
and $C_{\rm\scriptscriptstyle MS} = \ft53 - \gamma_{\rm E} + \ln 4 \pi$ in the MS
scheme, and $b = \ft{11}3 N_c - \ft43 T_F N_f$ is the first coefficient of the QCD
beta-function and $\alpha_s = \alpha_s (\mu^2) = 4 \pi /\left( b \ln \mu^2/
{\mit\Lambda}^2_{\overline{\rm\scriptscriptstyle MS}} \right)$, where the last
equality hold to one-loop order. The functions $F_M$ which enter the above structure
functions are combinations of $q$-flavor quark GPDs
\begin{equation}
F_\pi = F_u - F_d
\, , \qquad
F_\rho = Q_u F_u - Q_d F_d
\, , \qquad
F_\omega = Q_u F_u + Q_d F_d
\, ,
\end{equation}
where the quark charges are $Q_u = \ft23$ and $Q_d = - \ft13$. For $\pi^+$ and $V^0$
only the polarized $F = \{ \widetilde H, \widetilde E \}$ and, correspondingly,
unpolarized GPDs $F = \{ H, E \}$ enter the game. The quark coefficient function
(with resummed renormalon chains) for the $M = \pi^+$ has the form
\begin{eqnarray}
T_\pi (u, x, \xi; {\cal Q}^2)
\!\!\!&=&\!\!\! \frac{4 \pi C_F}{b}
\int_0^\infty \frac{d \tau}{\xi} \, {\rm e}^{- 4 \pi/(\alpha_s b) \tau}
\left( \frac{2 \mu^2 {\rm e}^C}{{\cal Q}^2} \right)^\tau
\nonumber\\
&\times&\!\!\!\Bigg\{
\frac{Q_u}{[ \bar u ( 1 - \frac{x}{\xi} - i 0 ) ]^{\tau + 1}}
-
\frac{Q_d}{[ u ( 1 + \frac{x}{\xi} - i 0 ) ]^{\tau + 1}}
\Bigg\}
\, ,
\end{eqnarray}
with $C_F = (N_c^2 - 1)/2N_c$. The coefficient function for neutral vector
mesons $M = V^0$, $T_V$, is obtained from this one by setting $Q_u, Q_d \to 1$
since the quark charges are included into flavor combinations of GPDs. Finally,
for completeness we present the leading-order gluon coefficient function
contributing to neutral vector meson production,
\begin{eqnarray*}
T_g (u, x, \xi) = \frac{\alpha_s}{\xi^2}
\frac{
4 T_F \sum_q Q_q
}{
u \bar u ( 1 - \frac{x}{\xi} - i 0 )( 1 + \frac{x}{\xi} - i 0 )
}
\, .
\end{eqnarray*}

If one absorbs the dependence on the momentum fraction into the argument
of the coupling, $\alpha_s ( \ft12 u (1 \pm \ft{x}{\xi}) {\cal Q}^2 {\rm e}^{- C})$,
one explicitly sees that the end-point regions produce divergences. Infrared
renormalons are caused by the end-point singularities [Feynman mechanism] in
exclusive amplitudes \cite{Aga96}, see also \cite{GodKiv98,KarSte01}. This can
be viewed as an estimate of the ambiguity in the resummation of higher-order
perturbative corrections or, taken to the extreme, as a model of higher-twist
contributions \cite{BenBraMag98}. Convolution of the coefficient function with
the distribution amplitude generates renormalon poles. For the asymptotic
distribution amplitude $\phi_{\rm asy} (u) = 6 u \bar u$ one gets two poles
$\tau = 1$ and $\tau = 2$, corresponding to ambiguities on the level of
${\cal Q}^{-2}$ and ${\cal Q}^{- 4}$ power corrections. Since the latter receives
extra contributions from higher order diagrams as well, we use only $\tau = 1$
pole for the estimates of the form of higher-twist corrections. Taking the
imaginary part (divided by $\pi$) arising from the contour deformation around
the renormalon poles as a measure of their magnitude, we get
\begin{equation}
\label{tilde-H}
\widetilde {\cal H}_\pi (\xi, \Delta^2; {\cal Q}^2)
=
\widetilde {\cal H}^{\rm\scriptscriptstyle PV}_\pi (\xi, \Delta^2; {\cal Q}^2)
+
\theta
\frac{{\mit\Lambda}^2_{\overline{\rm\scriptscriptstyle MS}}
\,
{\rm e}^{5/3}}{{\cal Q}^2}
\int_{- 1}^1 d x \, \Delta_{\widetilde H} (x, \xi) \widetilde H_\pi (x, \xi, \Delta^2)
\, ,
\end{equation}
where $\theta = \pm 1$ comes from the ambiguity to go around the renormalon pole
in the Borel plane. Here we have used the one-loop expression for the QCD coupling
constant and
\begin{equation}
\Delta_{\widetilde H} (x, \xi)
=
48 \frac{\pi C_F}{b \xi}
\Bigg\{
\frac{Q_u}{( 1 - \frac{x}{\xi} - i 0 )^2}
-
\frac{Q_d}{( 1 + \frac{x}{\xi} - i 0 )^2}
\Bigg\}
\, ,
\end{equation}
Since the GPD and its first derivative are continuous functions at $x = \pm \xi$
\cite{Rad99}, the second integral is well-defined. In the first term of
(\ref{tilde-H}) one uses the principal value prescription to go around the poles
in the Borel plane. For the $\widetilde E$ we use the pion-pole dominated form
\cite{GoePolVan01} and get
\begin{equation}
\widetilde {\cal E}_\pi (\xi, \Delta^2; {\cal Q}^2)
=
\widetilde {\cal E}^{\rm\scriptscriptstyle PV}_\pi (\xi, \Delta^2; {\cal Q}^2)
-
\theta
\frac{{\mit\Lambda}^2_{\overline{\rm\scriptscriptstyle MS}}
\,
{\rm e}^{5/3}}{{\cal Q}^2} \Delta_{\widetilde E} (\xi, \Delta^2; {\cal Q}^2)
\, ,
\end{equation}
where we have kept only the single and double poles at $\tau = 1$ in the
second term, so that
\begin{equation}
\Delta_{\widetilde E} (\xi, \Delta^2; {\cal Q}^2)
=
72 \frac{\pi C_F}{b \xi}
F_\pi (\Delta^2)
\left(
2
+
\ln \frac{{\mit\Lambda}^2_{\overline{\rm\scriptscriptstyle MS}}
\,
{\rm e}^{5/3}}{{\cal Q}^2}
\right)
\, .
\end{equation}
In the vicinity of the pion pole one can approximate $F_\pi (\Delta^2)
= 4 g_A M_N/(m_\pi^2 - \Delta^2)$.

\begin{figure}[t]
\mbox{
\begin{picture}(0,135)(209,0)
\put(0,3){\insertfig{4.45}{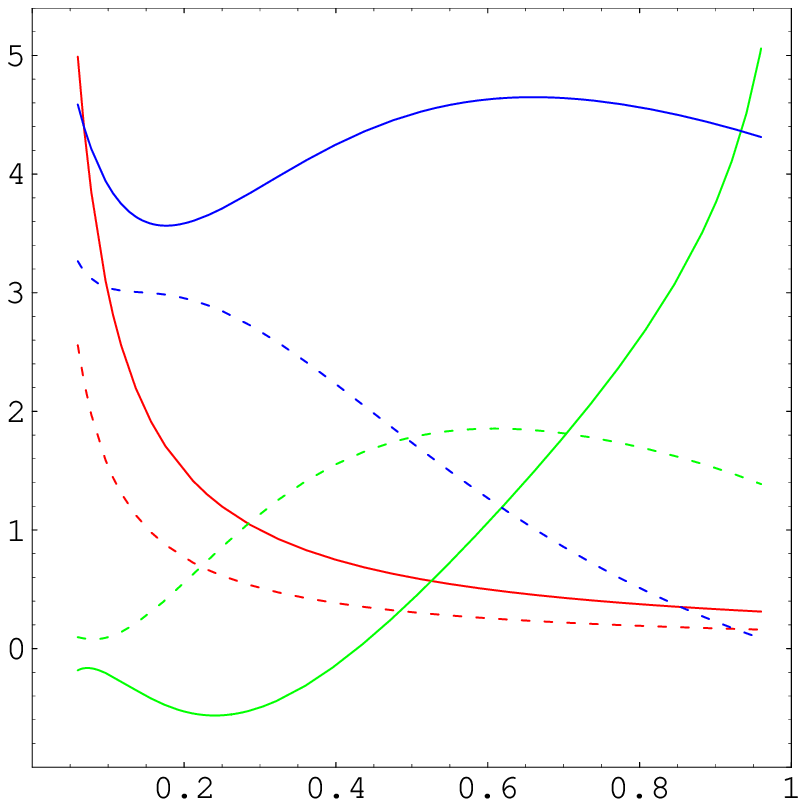}}
\put(23,87){${\scriptstyle 2}$}
\put(23,50.5){${\scriptstyle 3}$}
\put(23,24.5){${\scriptstyle 1}$}
\put(132,-4){\insertfig{4.88}{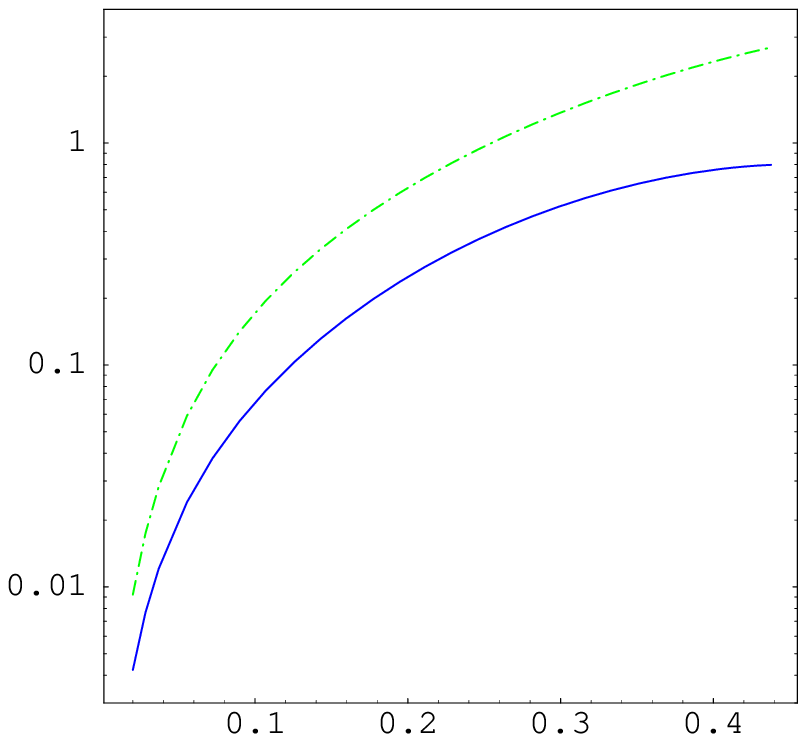}}
\put(170,20){$d \sigma_{\scriptscriptstyle L}^P/d|\Delta^2| \mbox{ vs. } \Bx$}
\put(285,0){\insertfig{4.65}{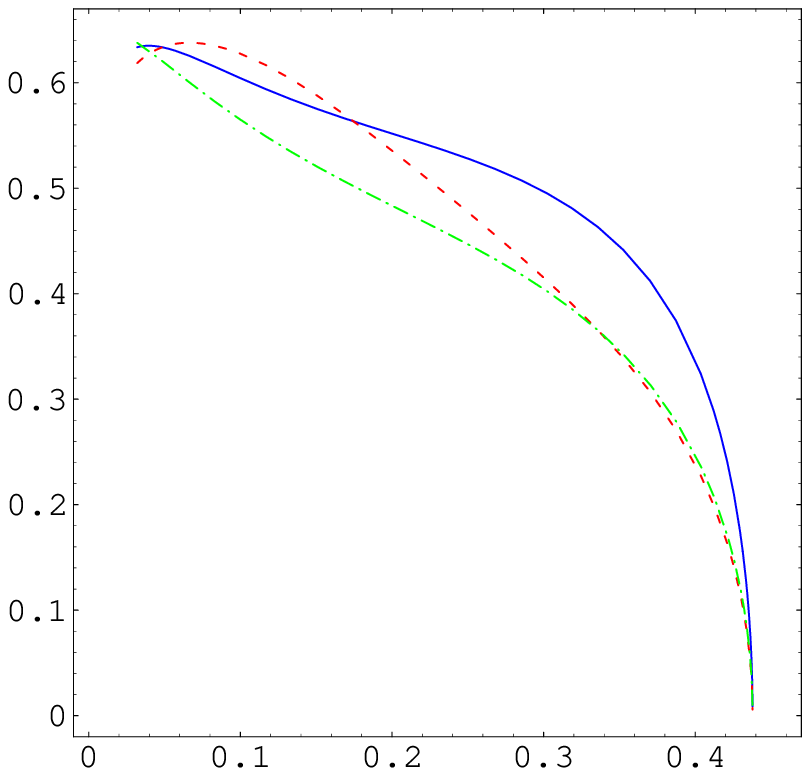}}
\put(330,20){${\cal A}^\perp_P \mbox{ vs. } \Bx$}
\end{picture}
}
\caption{\label{XsectionAndAsymmetry} Generalized structure functions (left)
in leading twist approximation (dashed) and including twist-four corrections
(solid) as a function of $x_{\rm B}$ for $\Delta^2 = - 0.3 \, {\rm GeV}^2$ and
${\cal Q}^2 = 10 \, {\rm GeV}^2$: (1) $\Re{\rm e} \widetilde {\cal H}$, (2)
$\Im{\rm m} \widetilde {\cal H}$, and (3) $10^{-2} \cdot \widetilde {\cal E}$.
The photoproduction cross section in units of nbarns (middle) without (solid)
and with (dash-dotted) power suppressed contributions for the same values of the
kinematical variables. The transverse spin asymmetry (right) at leading order
(solid) and with twist-four power effects taken into account for $\Delta^2
= - 0.3 \, {\rm GeV}^2$ and ${\cal Q}^2 = 4 \, {\rm GeV}^2$ (dashed) and
${\cal Q}^2 = 10 \, {\rm GeV}^2$ (dash-dotted). The maximal value of $x_{\rm B, max}$
is set by the kinematical constaint $|\Delta^2| > |\Delta^2_{\rm min}| =
M_N^2 \Bx^2/(1 - \Bx)$.}
\end{figure}

In estimates, shown in Fig.\ \ref{XsectionAndAsymmetry},  we relied on GPDs deduced
from an ansatz based on modeling the double distribution as a product \cite{Rad99}
${\mit\Delta} F (y, z, \Delta^2) = \pi (y, |z|) {\mit\Delta} f (y, \Delta^2)$ of
exclusive profile $\pi$ and an inclusive parton distribution augmented to have an
intrinsic momentum-transfer dependence ${\mit\Delta} f_q (y, \Delta^2) = \eta_q A_q
x^{a_q - \alpha'_q \Delta^2 (1 - x)}(1 - x)^{b_q} (1 + \gamma_q x + \rho_q \sqrt{x})$
with parameters fixed by the GSA forward densities \cite{GehSti96} in $\Delta^2 = 0$
limit and slopes $\alpha'_u = 1.15 \, {\rm GeV}^{-2}$, $\alpha'_d = 1.0 \,
{\rm GeV}^{-2}$ chosen to fit the dipole form of the axial form factor with the
effective mass $m_A^2 = 0.9 \, {\rm GeV}^2$. We give the cross section [at leading
order compatible with earlier estimates \cite{ManPilRad99,BelMul01}] and transverse
target-spin asymmetry ${\cal A}^\perp_P = (2 \sigma_P)/(\pi \sigma^\perp_P)$. In our
evaluations we set $\theta = 1$ and ${\mit\Lambda}_{\overline{\rm\scriptscriptstyle
MS}} = 280 \, {\rm MeV}$ for $N_f = 4$ and use the tree level result for
${\cal F}^{\rm\scriptscriptstyle PV} \to {\cal F}^{\rm\scriptscriptstyle LO}$. Note
however that in calculations of higher-twist corrections via renormalons in deeply
inelastic scattering in order to get the right magnitude of experimental data one
has to take a larger value $|\theta| \approx 2 - 3$ \cite{MauSteSchMan97}. The
extremely large power corrections to the absolute cross section of pion
leptoproduction are in qualitative agreement with the earlier consideration in
Ref.\ \cite{VanManSte98}. As we observe, however, the renormalon model of
higher-twist contributions affects in a marginal way the asymmetry and thus leads
to the apparent conclusion of {\sl the precocious scaling} in ratios of observables,
--- a fact pointed out previously in various circumstances
\cite{FraPobPolStr98,BelMul01,BelJiYua02}.

We would like to thank D. M\"uller and A.V. Radyushkin for conversations and INT
(Seattle) for its hospitality during the program ``GPDs and hard exclusive processes''.
This work was supported by the U.S.\ Department of Energy via grant DE-FG02-93ER-40762.


\end{document}